\documentstyle [12pt] {article}
\begin{document}
\setcounter{page}{1}
\def\theequation{\arabic{section}.\arabic{equation}}
\def\theequation{\thesection.\arabic{equation}}
\setcounter{section}{0}

\title{On the $D^{*+}_s \to D^+_s \pi^0$ decay in the Effective quark model with chiral $U(3)\times U(3)$ symmetry}

\author{A. N. Ivanov\thanks{E--mail: ivanov@kph.tuwien.ac.at, Institut f\"ur Kernphysik, Technische Universit\"at Wien, Wiedner Hauptsr. 8--10, A--1040 Vienna,  Austria }~\thanks{Permanent Address:
State Technical University, Department of Nuclear
Physics, 195251 St. Petersburg, Russian Federation}}

\date{}

\maketitle

\begin{center}
{\it Theory Division, CERN, CH--1211 Geneva 23, Switzerland}
\end{center}

\vskip1.0truecm
\begin{center}
\begin{abstract}
The partial width of the decay $D^{*+}_s \to D^+_s \pi^0$ going through the violation of isotopical symmetry is computed with respect to the partial width of the radiative decay $D^{*+}_s \to D^+_s \gamma$ in the effective quark model with chiral $U(3)\times U(3)$ symmetry incorporating Heavy quark effective theory (HQET) and Chiral perturbation theory at the quark level (CHPT)$_q$. We investigate  a sensitivity of the  ratio $R_0 = \Gamma(D^{*+}_s \to D^+_s \pi^0)/\Gamma(D^{*+}_s \to D^+_s \gamma)$ to current $s$--quark mass corrections and find a strong dependence.
\end{abstract}
\end{center}
\vspace{0.2in}

\begin{center}
PACS number(s): 13.30.Ce, 12.39.Ki, 14.20.Lq, 14.20.Mr
\end{center}
\newpage

\section{Introduction}
\setcounter{equation}{0}

A formulation of Heavy quark effective theory (HQET) [1--3] based on the infinite limit of heavy quark masses gave a substantial impact to the development of a quantitative non--perturbative approach to physics of heavy--light quark states [4]. In HQET heavy quarks are static colour sources and light constituents couple to them via the exchange of soft--gluons. Since  at present low--energy QCD is not self--consistently completed, different phenomenological approaches motivated by QCD are still actual. As has been shown in [5--9], the low--energy interactions of heavy quarks with light constituents can be quantitatively described within Chiral perturbation theory at the quark level (CHPT)$_q$ [10] with linear realization of chiral $U(3)\times U(3)$ symmetry. Supplying HQET by (CHPT)$_q$ we arrive at an effective quark model taking into account all requirements of chiral symmetry. In such an effective model with chiral $U(3)\times U(3)$ symmetry we can investigate  mass spectra of heavy--light mesons, their coupling constants and processes of heavy--light hadron decays for which the kinetic energies of the particles in the final state do not exceed the scale of spontaneous breaking of chiral symmetry (SB$\chi$S) $\Lambda_{\chi}\sim 1\,{\rm GeV}$.

This paper is to apply the effective quark model with chiral $U(3)\times U(3)$ symmetry to the computation of the $D^{*+}_s \to D^+_s \pi^0$ decay  which comes through the violation of isotopical symmetry. The partial width $\Gamma(D^{*+}_s \to D^+_s \pi^0)$ of the $D^{*+}_s \to D^+_s \pi^0$ decay has been measured recently with respect to the partial width $\Gamma(D^{*+}_s \to D^+_s \gamma)$ of the radiative decay $D^{*+}_s \to D^+_s \gamma$ [11]:
\begin{eqnarray}\label{label1.1}
R^{\exp}_0 = \frac{\Gamma(D^{*+}_s \to D^+_s \pi^0)}{\Gamma(D^{*+}_s \to D^+_s \gamma)} = 0.062\pm 0.027.
\end{eqnarray}
The first theoretical consideration of the $D^{*+}_s \to D^+_s \pi^0$ decay has been carried out by Cho and Wise [12], who used the $SU(3)$ limit for the computation of the partial width of the radiative decay $D^{*+}_s \to D^+_s \gamma$ equating it to the partial width of the decay  $D^{*+} \to D^+ \gamma$, i.e. $\Gamma(D^{*+}_s \to D^+_s \gamma) = \Gamma(D^{*+} \to D^+  \gamma)$. In such an approximation they gave the ratio $R_0$ ranging $0.01\div 0.10$ [12,13].

However, as has been shown in [8] the partial width $\Gamma(D^{*+}_s \to D^+_s \gamma)$ is very sensitive to current $s$--quark mass corrections. Below we analyze a sensitivity of the ratio $R_0$ to current $s$--quark mass corrections and find a strong dependence on the current $s$--quark mass.

\section{Partial width of the $D^{*+}_s \to D^+_s \pi^0$ decay}
\setcounter{equation}{0}

Following Cho and Wise [12] we would take into account only the effect of the isotopical spin violation induced by the mass difference of current $u$-- and $d$--quarks described by the Gasser--Leutwyler Lagrangian [14]
\begin{eqnarray}\label{label2.1}
{\cal L}^{\Delta I=1}_{\rm QCD}(x) = \frac{1}{2}\,(m_{0 d} - m_{0 u})\,[\bar{u}(x)u(x) - \bar{d}(x)d(x)],
\end{eqnarray}
where $m_{0 u} = 4\,{\rm MeV}$ and $m_{0 d}=7\,{\rm MeV}$ determined at the normalization scale $\mu =1\,{\rm GeV}$ [13], $u(x)$ and $d(x)$ are the operators of the current $u$-- and $d$--quark fields.

The amplitude of the decay $D^{*+}_s \to D^+_s \pi^0$ can be given in the form
\begin{eqnarray}\label{label2.2}
M(D^{*+}_s(Q) \to D^+_s(p) \pi^0(q)) = g_{D^{*+}_s D^+_s \pi^0}\,e(Q)\cdot q,
\end{eqnarray}
where $e^{\mu}(Q)$ is the 4--vector of the $D^{*+}_s$--meson polarization and $q^{\mu}$ is a 4--momentum of the $\pi^0$--meson.

For the computation of the $g_{D^{*+}_s D^+_s \pi^0}$ coupling constant we take into account the contributions of the $\eta(550)$ and $\eta^{\prime}(960)$ meson intermediate states. In this case the $g_{D^{*+}_s D^+_s \pi^0}$ coupling constant reads
\begin{eqnarray}\label{label2.3}
&&g_{D^{*+}_s D^+_s \pi^0} =\nonumber\\
&&=- g_{D^{*+}_s D^+_s \eta_s}\,\frac{1}{4}\,\Bigg(\frac{m_{0 d} - m_{0 u}}{M^2_{\eta} - M^2_{\pi^0}}- \frac{m_{0 d} - m_{0 u}}{M^2_{\eta^{\prime}} - M^2_{\pi^0}}\Bigg) <\pi^0|[\bar{u}(0)u(0) - \bar{d}(0)d(0)]|\eta_{\rm N}> \sin 2\bar{\theta}\nonumber\\
&&= - g_{D^{*+}_s D^+_s \eta_s}\,\frac{1}{4}\,\frac{m_{0 d} - m_{0 u}}{M^2_{\eta} - M^2_{\pi^0}}\frac{M^2_{\eta^{\prime}} - M^2_{\eta}}{M^2_{\eta^{\prime}} - M^2_{\pi^0}} <\pi^0|[\bar{u}(0)u(0) - \bar{d}(0)d(0)]|\eta_{\rm N}> \sin 2\bar{\theta}
\end{eqnarray}
where $\eta_{\rm N}$ and $\eta_s$ are the pseudoscalar isotopical singlets with the quark structure $\eta_{\rm N}=(\bar{u}u + \bar{d} d)/\sqrt{2}$ and $\eta_s = \bar{s}s$, respectively. The states $\eta_{\rm N}$ and $\eta_s$ are mixed in the observed $\eta$ and $\eta^{\prime}$ mesons with the mixing angle $\bar{\theta} = \theta_0 - \theta_P$ [14]
\begin{eqnarray}\label{label2.4}
\eta &=&\eta_N\,\sin\bar{\theta} - \eta_s\,\cos \bar{\theta},\nonumber\\
\eta^{\prime} &=&\eta_N\,\cos\bar{\theta} + \eta_s\,\sin \bar{\theta},
\end{eqnarray}
where ${\rm tg}\,\theta_0 = 1/\sqrt{2}$ and $\theta_P$ is the singlet--octet mixing angle [11]. 

Then $g_{D^{*+}_s D^+_s \eta_s}$ is the coupling constant of the strong $D^{*+}_s D^+_s \eta_s$--interaction. In the chiral limit $g_{D^{*+}_s D^+_s \eta_s}$ amounts to the coupling constant $g_{D^* D\pi}$ of the strong $D^* D\pi$--interaction, i.e. $g_{D^{*+}_s D^+_s \eta_s} = g_{D^* D\pi}$. The coupling constant $g_{D^* D\pi}$ has been computed within HQET supplemented by (CHPT)$_q$ in Ref. [6,9] (see also [7]). Therefore, in the chiral limit we have
\begin{eqnarray}\label{label2.5}
g_{D^{*+}_s D^+_s \eta_s} = g_{D^* D\pi} = \frac{4\pi}{\sqrt{N}}\frac{\sqrt{M_D M_{D^*}}}{\bar{v}^{\prime}}{\ell n}\Bigg(\frac{\bar{v}^{\prime}}{4m}\Bigg) = 5.3,
\end{eqnarray}
where $N=3$ is the number of quark colours, $M_D=1.86\,{\rm GeV}$ and  $M_{D^*} = 2.00\,{\rm GeV}$ are  the masses of $D$ and $D^*$ mesons in the chiral limit [7], $m = 0.33\,{\rm GeV}$ is the light constituent quark mass calculated in the chiral limit [10], and $\bar{v}^{\prime}= 4\,\Lambda = 2.68\,{\rm GeV}$ is the cut--off in Euclidean 3--momentum space connected with the SB$\chi$S scale in (CHPT)$_q$ by the relation $\Lambda = \Lambda_{\chi}/\sqrt{2}= 0.66\,{\rm GeV}$ [5--9] at $\Lambda_{\chi}=0.94\,{\rm GeV}$ [10].

One can show that in the suggested approach current $s$--quark mass corrections to the coupling constant $g_{D^{*+}_s D^+_s \eta_s}$ appear only due to chiral corrections to the masses of $D^+_s$ and $D^{*+}_s$ mesons [7]:
\begin{eqnarray}\label{label2.6}
\hspace{-0.3in}M_{D^+_s} = M_D \Bigg[1 + \frac{m_{0 s}}{2 m} \frac{\bar{v}}{\bar{v}^{\prime}}\,{\ell n}\Bigg(\frac{\bar{v}^{\prime}}{4m}\Bigg)\Bigg],M_{D^{*+}_s} = M_{D^*} \Bigg[1 + \frac{m_{0 s}}{2 m}\,\frac{\bar{v}}{\bar{v}^{\prime}} {\ell n}\Bigg(\frac{\bar{v}^{\prime}}{4m}\Bigg)\Bigg],
\end{eqnarray}
where $m_{0 s} =135\,{\rm MeV}$ is the mass of the current $s$--quark [14,7,10] and $\bar{v} = -<\!\bar{q}q\!>/F^2_0 = 1.92\,{\rm GeV}$ with $F_0 = 92\,{\rm MeV}$ the leptonic coupling constant of light pseudoscalar mesons calculated in the chiral limit [10].

Replacing in Eq.(\ref{label2.5}) the masses $M_D$ and $M_{D^*}$ calculated in the chiral limit by $M_{D^+_s}$ and $M_{D^{*+}_s}$ given by Eq.(\ref{label2.6}), we obtain the coupling constant $g_{D^{*+}_s D^+_s \eta_s}$ up to first order in current $s$--quark mass expansion
\begin{eqnarray}\label{label2.7}
g_{D^{*+}_s D^+_s \eta_s} = g_{D^* D\pi}\,\Bigg[1 + \frac{m_{0 s}}{2 m}\,\frac{\bar{v}}{\bar{v}^{\prime}}\,{\ell n}\Bigg(\frac{\bar{v}^{\prime}}{4m}\Bigg)\Bigg] = 5.9.
\end{eqnarray}
The matrix element $<\pi^0|[\bar{u}(0)u(0) - \bar{d}(0)d(0)]|\eta_{\rm N}>$ has been computed in Ref. [10] and reads
\begin{eqnarray}\label{label2.8}
&&<\pi^0|[\bar{u}(0)u(0) - \bar{d}(0)d(0)]|\eta_{\rm N}>=<\pi^0|[\bar{u}(0)u(0) + \bar{d}(0)d(0)]|\pi^0> =\nonumber\\
&&= <\pi^+|[\bar{u}(0)u(0) + \bar{d}(0)d(0)]|\pi^+> = 2\,\bar{v},
\end{eqnarray}
which is in accordance with the Gell--Mann--Oakes--Renner theorem [15]. The relations Eq.(\ref{label2.8}) are caused by isotopical invariance.

For the computation of $g_{D^{*+}_s D^+_s \pi^0}$ we use the experimental values of the meson masses $M_{\eta} = 547\,{\rm MeV}$, $M_{\eta^{\prime}} = 958\,{\rm MeV}$ and the mixing angle $\theta_{\rm P} = - 20^0$ [11].

Collecting the contributions we arrive at the following value of the $g_{D^{*+}_s D^+_s \pi^0}$ constant
\begin{eqnarray}\label{label2.9}
g_{D^{*+}_s D^+_s \pi^0} = - g_{D^{*+}_s D^+_s \eta_s}\,\frac{1}{2}\,\frac{(m_{0 d} - m_{0 u})\bar{v}}{M^2_{\eta} - M^2_{\pi^0}}\,\frac{M^2_{\eta^{\prime}} - M^2_{\eta}}{M^2_{\eta^{\prime}} - M^2_{\pi^0}}\,\sin 2\bar{\theta} = - 0.04,
\end{eqnarray}
where we have set $M_{\pi^0} = 135\,{\rm MeV}$ [11]. The partial width $\Gamma(D^{*+}_s \to D^+_s \pi^0)$ is then given by
\begin{eqnarray}\label{label2.10}
\Gamma(D^{*+}_s \to D^+_s \pi^0) = \frac{g^2_{D^{*+}_s D^+_s \pi^0}}{24\pi}\frac{|\vec{q}\,|^3}{M^2_{D^{*+}_s}} = 5.4\times 10^{-10}\,{\rm GeV}.
\end{eqnarray}
The relative momentum $|\vec{q}\,| = 48.4\,{\rm MeV}$ has been computed at $M^2_{D^{*+}_s} = 2113\,{\rm MeV}$, $M^2_{D^+_s} = 1969\,{\rm MeV}$ and $M_{\pi^0} = 135\,{\rm MeV}$ [11].

In order to compute the ratio $R_0$ we have to know the value of the partial width $\Gamma(D^{*+}_s \to D^+_s \gamma)$ of the radiative decay $D^{*+}_s \to D^+_s \gamma$. The former reads [8]
\begin{eqnarray}\label{label2.11}
\Gamma(D^{*+}_s \to D^+_s \gamma) = \frac{\alpha}{3}\,g^2_{D^{*+}_s D^+_s \gamma}\Bigg(\frac{M^2_{D^{*+}_s} - M^2_{D^+_s}}{2 M_{D^{*+}_s}}\Bigg)^{\!3} = 5.3\times 10^{-8}\,{\rm GeV},
\end{eqnarray}
where $\alpha =1/137$ is the fine structure constant and $g_{D^{*+}_s D^+_s \gamma}$ is defined [8]
\begin{eqnarray}\label{label2.12}
\hspace{-0.3in}g_{D^{*+}_s D^+_s \gamma}&=&\sqrt{\frac{M_D}{M_{D^*}}}\Bigg[-\frac{2}{3}\frac{1}{\bar{v}^{\prime}}\,{\ell n}\Bigg(\frac{\bar{v}^{\prime}}{4m}\Bigg)\Bigg(1-\frac{m}{M_c}\Bigg) + \frac{1}{6}\frac{1}{M_c} - \frac{m_{0 s}}{M_c}\frac{1}{3m}\frac{\bar{v}}{\bar{v}^{\prime}}\,{\ell n}\Bigg(\frac{\bar{v}^{\prime}}{4m}\Bigg)\Bigg]\nonumber\\
&=&-0.09 \,{\rm GeV}^{-1},
\end{eqnarray}
where $M_c$ is the $c$--quark mass, $M_D = M_c =1.86\,{\rm GeV}$ and $M_{D^*}=2.00\,{\rm GeV}$ [5--9]. Since the chiral correction in current $s$--quark mass expansion enters to the coupling constant as the ratio $m_{0 s}/M_c$, the next--to--leading order corrections in large $M_c$ expansion have been taken into account too [8].

For the ratio $R_0(m_{0 s})$ calculated to next--to--leading order in current $s$--quark mass expansion we obtain 
\begin{eqnarray}\label{label2.13}
R_0(m_{0 s})= \frac{\Gamma(D^{*+}_s \to D^+_s \pi^0)}{\Gamma(D^{*+}_s \to D^+_s \gamma)} = 0.010.
\end{eqnarray}
In turn, at leading order in current $s$--quark mass expansion we get
\begin{eqnarray}\label{label2.14}
R_0(0) = \frac{\Gamma(D^{*+}_s \to D^+_s \pi^0)}{\Gamma(D^{*+}_s \to D^+_s \gamma)} = 0.028.
\end{eqnarray}
Thus, we have shown that the ratio $R_0$ is very sensitive to the current $s$--quark mass corrections.

Our results both Eq.(\ref{label2.13}) and Eq.(\ref{label2.14}) agree well with the constraint by Cho and Wise, i.e. $R_0 = 0.01\div 0.10$. However, the theoretical ratios disagree with the experimental data  $R^{\exp}_0 = 0.062\pm 0.027$ [11].

\section{Conclusion}

We have shown that the ratio $R_0(m_{0 s})$ calculated in the effective quark model with linear realization of chiral $U(3)\times U(3)$ symmetry, incorporating HQET and (CHPT)$_q$, is very sensitive to current $s$--quark mass corrections. We have found that $R_0(0)/R_0(m_{0 s}) = 2.8$, where $R_0(0) = 0.028$ and $R_0(m_{0 s}) = 0.010$ are the ratios calculated at leading and to next--to--leading order in current $s$--quark mass expansion, respectively. For the both cases the theoretical values of the ratio $R_0(0) = 0.028$ and $R_0(m_{0 s}) = 0.010$ satisfy the constraint by Cho and Wise, i.e., $R_0 = 0.01 \div 0.10$, but disagree with the experimental data $R^{\exp}_0 = 0.062\pm 0.027$ [11]. Of course, the ratio $R_0$ is proportional to $(m_{0 d} - m_{0 u})^2$ as
\begin{eqnarray*}
R_0  = (0.028\div 0.010)\times \Bigg(\frac{m_{0 d} - m_{0 u}}{3}\Bigg)^2,
\end{eqnarray*}
and by tuning this difference one can fit the experimental data. For example, starting with $m_{0 d} - m_{0 u} \ge 5.6\,{\rm MeV}$ instead of $m_{0 d} - m_{0 u} =3\,{\rm MeV}$ the theoretical ratio $R_0$ would be made in agreement with the experimental data. However, in our approach the mass difference $m_{0 d} - m_{0 u}$ is strictly fixed to be equal to $m_{0 d} - m_{0 u} =3\,{\rm MeV}$ through the mass difference of the $K^+$ and $K^0$ mesons [10]. Hence, by taking into account current $s$--quark mass corrections we cannot predict more than $R_0(m_{0 s}) = 0.010$.

In this connection we would emphasize that chiral corrections in current $s$--quark mass expansion have been calculated in the tree--meson approximation. A strong dependence of the ratio $R_0(m_{0 s})$ on $m_{0 s}$ makes sense to take into account  one--meson loop chiral corrections which should lead  to the appearance of chiral logarithms like $m_{0 s}\,{\ell n}(m_{0 s})$. The account for these chiral logarithms goes beyond the scope of this paper and would be the matter of our further investigations.

\section{Acknowledgment}

The author is grateful to TH--Division for kindful hospitality during a stay at CERN, where this work has been done, and to Prof. M. Neubert for reading  of the manuscript and numerous useful comments. The discussions with Prof. S. Narison and Prof. N. I. Troitskaya are appreciated.

\end{document}